\documentclass[twocolumn]{aastex631}

\shorttitle{Born to be wide}
\shortauthors{Rozner \& Perets}

\usepackage{graphicx}	
\usepackage{amsmath}	
\usepackage{amssymb}	
\usepackage{color}
\usepackage{float}
\usepackage{ulem}
\usepackage{soul}
\usepackage{lipsum}
\usepackage[T1]{fontenc}

\usepackage{xcolor}

\DeclareRobustCommand{\VAN}[3]{#2}
\let\VANthebibliography\thebibliography
\def\thebibliography{\DeclareRobustCommand{\VAN}[3]{##3}\VANthebibliography}

\usepackage{graphicx}	
\usepackage{amsmath}	
\usepackage{braket}

\begin{document}


\title{Born to be wide: the distribution of wide binaries in the field and soft binaries in clusters}


\email{morozner@campus.technion.ac.il}

\author[0000-0002-2728-0132]{Mor Rozner}
\affiliation{Technion - Israel Institute of Technology, Haifa, 3200002, Israel}


\author[0000-0002-5004-199X]{Hagai B. Perets}
\affiliation{Technion - Israel Institute of Technology, Haifa, 3200002, Israel}
\affiliation{Department of Natural Sciences, The Open University of Israel, 1 University Road, PO Box 808, Raanana 4353701, Israel}

\begin{abstract}
Most stars, binaries, and higher multiplicity systems are thought to form in stellar clusters and associations, which later dissociate. Very wide binaries can be easily disrupted in clusters due to dynamical evaporation (soft binaries) and/or due to tidal disruption by the gravitational potential of the cluster. Nevertheless, wide binaries are quite frequent in the field, where they can sometimes play a key role in the formation of compact binaries, and serve as tools to study key physical processes. 
Here we use \emph{analytic} tools to study the dynamical formation of soft binaries in clusters, and their survival as field binaries following cluster dispersion. We derive the expected properties of very wide binaries both in clusters and in the field. We analytically derive their detailed distributions, including wide-binary fraction as a function of mass in different cluster environments, binaries mass functions and mass ratios, and the distribution of their orbital properties. We show that our calculations agree well on most aspects with the results of N-body simulations, but show some different binary-fraction dependence on the cluster mass.  
We find that the overall fraction of wide binaries scales as $\propto N_\star^{-1}$ where $N_\star$ is the size of the cluster, even for non-equal mass stars. More massive stars are more likely to capture wide companions, with most stars above five solar mass likely to capture at least one stellar companion, and triples formation is found to be frequent. 
  
\end{abstract}


\section{Introduction} 

Binary and higher multiplicity stellar systems are quite frequent (e.g. \citealp{DuquennoyMayor1991,Raghavan2010,Sana2012,DucheneKraus2013,MoeStefano2017}), and play a key role 
in the dynamics and evolution of stellar systems.  Most stars, even field stars,  are thought to have formed in stellar clusters and associations \citep{LadaLada2003}, and later released to the field once their host clusters dispersed. The early dynamics of binaries can therefore be significantly altered by their interactions with other stars, and/or due to the overall potential of their host clusters.  
In dense environments, 
binaries could be divided into two groups, based on the energy relative to the mean energy of their background: dynamically-hard binaries ($|E_{\rm bin}|\gtrsim\bar m \sigma^2$) and dynamically-soft binaries ($|E_{\rm bin}|\lesssim\bar m \sigma^2$).
The evolution of soft and hard binaries differ qualitatively in such environments. While hard binaries become, on average, harder, due to interactions with other stars (more compact, shorter periods, or more general i.e. larger absolute binding energy, accounting for exchanges),  soft binaries become softer \citep{Heggie1975,Hills1975_1}. It should be noted that hard binaries could go through exchanges when they encounter a third perturber. See a revised version, accounting for external cutoffs, of Heggie's law in \cite{GinatPerets2021}.

In the field, the low binding energy of 
wide binaries ($a\gtrsim 10^3 \ \rm{AU}$) make them sensitive to even far flyby perturbations and other gravitational perturbations. This makes them an important tool to probe the galactic potential, MACHOs and primordial black-holes (e.g. \citealp{BahcallHutTremaine1985,ChanameGould2004,Quinn2009,Blas2017,AxionsBinaries_Rozner2020}). In addition, flyby and galactic tidal perturbations  sometimes excite their eccentricities to extreme values, allowing the wide binary (and wide triple) components to closely interact through tidal, gravitational-wave of direct collisional interactions, giving rise to the formation of compact interacting binaries and/or merger products \citep{KaibRaymond2014,MichaelyPerets2019,MichaelyPerets2020,GrishinPerets2022,MichaelyNaoz2022}.  Wide binaries could also constrain star formation (e.g. \citealp{Larson2001}).

Given their important roles, it is important to understand the origins of wide binaries.
Soft, wide binaries, can be dynamically formed in dense environments where perturbation by other stars can change their velocities, and give rise to random binding of two stars, with appropriate relative velocities and separations.   
It was suggested that the observed wide binaries in the field are surviving soft binaries that formed in the birth-cluster/association, after these clusters dispersed\footnote{Other suggestions of dynamical formation in the field are unlikely \citep{GoodmanHut93}.}. These are essentially the most recently formed soft binaries in the clusters which then survived as their host cluster dissolved/formed during the dissolution
\citep{Kouwenhoven_eatl2010,MoeckelBate2010,MoeckelClarke2011,PeretsKouwenhoven2012}.  
Other formation channels are  the formation of primordial wide binaries directly in the early star-formation phases in a gaseous environment, e.g. by fragmentation (e.g. \citealp{Duchene2004,Offner2010}), gas-induced captures \citep{Tagawa2020, Rozner_gas2022,SiyaoLai2023} and dynamical unfolding of compact triples \citep{ReipurthMikkola2012}.  These originally formed wide binaries might not survive, due to perturbations, unless hardened by gas-induced inspiral, making them close binaries, or if they formed at late stages just before the cluster dispersal, at which point gas might not be available. Although the formation process of wide binaries is still unknown, chemical similarities were observed between binaries components, which indicate that the components of wide binaries were born together and were not formed due to random pairing (e.g. \citealp{Andrews2018,Hawkins2020}). 

In the field, wide binaries may still experience infrequent flyby perturbations from field stars, and the widest ones could eventually be disrupted due to the tidal field of the Galaxy \citep{Ambartsumian1937,Chandrasekhar1944,Yabushita1966,Heggie1975,King1977,Heggie1977,RettererKing1982, BahcallHutTremaine1985,JiangTremaine2010},  or potentially be excited to high eccentricities leading to strong peri-center interactions or collisions between the binary components due to Galactic tide induced secular evolution \citep{HeislerTremaine1986,BonsorVeras2015,HamiltonRafikov2019a,HamiltonRafikov2019b}.

Explaining the formation and survival of wide binaries is challenging, due to their wide separation that could in principle reach the size of the core of a young cluster, and their sensitivity to perturbations, which can destroy them.

Nevertheless, wide binaries are quite frequent \citealp[e.g][]{El-badryRix2018,ElBadry2019,ElBadry2021}.  Here we focus on the dynamical formation channel of soft binaries and their survival and provide the first analytic study of their detailed properties.

In this paper, we derive analytically the distributions of wide binaries in clusters and in the field, for general mass functions. We then compare our results with N-body simulations and observations.

In section \ref{sec: wide binaries distribution}, we derive analytically  the distributions and the overall fractions of wide/soft binaries. In section \ref{sec:results} we present the results from our Monte Carlo simulations, based on the analytical derivations. In section \ref{sec:discussion} we discuss our results and future implications. In section \ref{sec:summary}, we summarize and conclude.

\section{The Distribution of soft/Wide Binaries}\label{sec: wide binaries distribution}

To enable the dynamical formation of a binary, two unbound stars need to be perturbed and change their relative velocities, to become bound. 
The conditions under which a bound binary is formed in this case are the following: (I) the binary separation should be sufficiently small such that the binary should survive the tidal radius of the cluster (for soft binaries in clusters) or the Galactic tidal radius (for wide binaries in the field). (II) The two components should become bound, i.e. their relative velocity should be smaller than the escape velocity at their instantaneous distance. 

In our calculation, we define and characterize the available parameter space for wide binaries.
 
\subsection{Clusters in Virial equilibrium -- soft binaries in clusters }\label{subsec:virial equilibrium}
The distribution of energetically-soft binaries could be derived under the assumption of thermal equilibrium, using mechanical statistics methods, since their weak binding  enables them to reach equilibrium faster than a whole stellar system, which in general has no maximum-entropy state \citep{GoodmanHut93,BinneyTremaine2010}.

Consider a cluster with $N$ stars with masses $\left\{m_i\right\}_{i=1}^N$, contained in a volume $V$. The temperature of the cluster is defined by $\beta^{-1}:= k_BT=\bar m \sigma^2$, where $\bar m$ is the mean mass in the cluster
 and $\sigma$ is its velocity dispersion.
 When planets are considered as well, the mass is still taken to be the mean mass of the stars, as the planets have a negligible effect on the cluster structure.
 Binaries with larger energies than the cluster temperature are hard binaries, while those with lower energies are soft. 

The distribution of a binary constitutes stars of masses $m_1$ and $m_2$, in thermal equilibrium, is given by Boltzmann distribution, i.e.  $f_{1,2}(\bf{w_1, w_2})\propto e^{-\beta \mathcal H}$ where $\bf w_i$ are the coordinates in the 6-dimensional phase space and $\mathcal H$ is the two-body corresponding Hamiltonian, 

\begin{align}
\mathcal H = \frac{1}{2}m_1v_1^2 +\frac{1}{2}m_2 v_2^2-\frac{Gm_1m_2}{|x_1-x_2|}
\end{align}
The number density of binaries with internal energy $\tilde E$ is then given by 

\footnotesize
\begin{align}
n_{\rm eq}^{1,2}(\tilde E)=\frac{N_1N_2}{V}\int f_{1,2}(w_1,w_2) \delta\left(\tilde E(w_1,w_2)-\tilde E\right)d^6 w_1 d^6w_2
\end{align}
\normalsize

\noindent 
where $N_i$ is the number of stars from the i-th species. 
The total number of binaries is 

\begin{align}\label{eq:neqE}
n_{\rm eq}^{1,2}(\tilde E)= 
\frac{G^3 \rho_1 \rho_2 \pi^{3/2}(m_1m_2)^2}{8\mu^{3/2}\sigma^3 |\tilde E|^{5/2}}\left(\frac{\bar m}{M}\right)^{3/2}e^{|\tilde E|/\bar m \sigma^2}
\end{align}

\noindent 
where $\rho_i=n_im_i$ are the densities of the different species $\mu_{12}=m_1m_2/(m_1+m_2)$ is the reduced mass of a binary and $M=m_1+m_2$ is the total mass of the binary. It should be noted that here we consider a uniform stellar density, as a simplifying assumption. However, in a more realistic model of clusters, other density profiles should be taken into consideration, which could modify the outcomes.

The number density for a given semi-major axis $a$ is 

\begin{align}\label{eq:n(a)}
n^{1,2}_{\rm eq}(a)=
\frac{G^{3/2}n_1n_2 \pi^{3/2}\bar m^{3/2}}{2^{3/2}\sigma^3}a^{1/2}e^{|\tilde E(a)|/\bar m \sigma^2}
\end{align}

For the simplified case of clusters composed of single-mass stars, there is an agreement with the analytical expressions derived in \cite{GoodmanHut93,BinneyTremaine2010}.
The total number of soft/wide binaries could be derived via the integration of eq. \ref{eq:n(a)}, where the integral is dominated by the lower boundary, i.e. the minimal energy allowed for a soft/wide binary in the relevant context. While the separation of soft binaries in clusters could not exceed the Hill radius of the cluster, immediately after a cluster dispersion, leftover binaries from the cluster would become part of the field population and could survive with separations as large as the galactic tidal radius. 
Henceforth, we will split our analysis for soft binaries into two regimes: binaries in clusters and wide binaries in the field (i.e from dispersed clusters). Each regime dictates different regimes of integration and hence different distributions.  

The maximal energy for a soft binary is determined by the transition between hard and soft binaries and given by $\bar m \sigma^2$. The upper limit for the separation is 
the Hill radius defined by $R_{\rm {Hill,c}} = \left((m_1+m_2)/M_{\rm cluster}\right)^{1/3}R_{\rm cluster}$ where $M_{\rm cluster}$ and $R_{\rm cluster}$ are the mass and radius of the cluster correspondingly. 
The galactic tidal field sets a larger cutoff separation, given by $R_{\rm{Hill,g}}\sim 1.7((m_1+m_2)/2M_\odot) \ \rm{pc}$ \citep{JiangTremaine2010}.

We can integrate eq. \ref{eq:neqE} to obtain the total number density of soft binaries taking into account the contributions from the whole energy range. For simplicity, we approximate the exponential as unity.

We assume a uniform density of soft binaries inside the cluster, and approximate $N_{\rm soft}^{1,2}\approx n_{\rm{eq}}^{1,2}V$ and $\rho_i=m_i N_i/V$.

\small
\begin{align}\label{eq:Fraction}
&N^{1,2}_{\rm soft}\approx  R_{\rm cluster}^3n_{\rm eq}^{1,2}\approx \\
\nonumber
&\approx
\frac{N_1N_2\pi^{3/2}}{12M_{\rm cl}^2}\bar m^{3/2}M^{1/2}-\frac{N_1N_2 \pi^{3/2}}{12}\left(\frac{m_1m_2}{M_{\rm cl}^2}\right)^{3/2}
\end{align}
\normalsize

For any given lower energy cutoff $|\tilde E_{\rm min}|$, or equivalently a maximal separation $R_{\rm cut}$,  the number of binaries will be given by 

\footnotesize
\begin{align}
&N_{\rm bin}^{1,2}(R_{\rm cut})
\approx \\
\nonumber
&\approx
\frac{(2\pi)^{3/2}G^{3/2}n_1n_2 \bar m^{3/2}}{12\sigma^3}R_{\rm cluster}^3R_{\rm cut}^{3/2}-\frac{N_1N_2 \pi^{3/2}}{12}\left(\frac{m_1m_2}{M_{\rm cl}^2}\right)^{3/2}
\end{align}

\normalsize

\noindent
Where for soft binaries, $R_{\rm cut}=R_{\rm Hill}$. For a general mass distribution $\xi(m)$, 

\begin{align}
N_{\rm bin}^{\rm total}(R_{\rm cut}) = \int N^{1,2}_{\rm bin}(R_{\rm cut})\xi(m_1)\xi(m_2)dm_1dm_2
\end{align}

\noindent
For single mass clusters, it can be seen that the number of existing systems at any given point is of order unity, as expected. 

It should be noted that while our derivation was based on the assumption of thermal equilibrium, the distribution we obtained is in principle valid for a more general case, in which the only assumption is that the distribution of the background stars is Maxwellian, which is less demanding. Then, the final distribution is guaranteed based on the principle of detailed balance \citep{Heggie1975,BinneyTremaine2010}.

\subsection{Expanding clusters -- wide binaries in the field}
\label{subsec:expanding}

In later stages of cluster evolution, the cluster tends to dissolve, due to the expulsion of primordial gas and dynamical processes. Hence, a more realistic description of it should include expansion.  
We can extend our derivation to describe an expanding cluster. The Virial ratio is defined as the ratio between the kinetic and potential energies of the systems, i.e. $Q=-K/U$. Clusters with $Q=1/2$ are in Virial equilibrium, $Q<1/2$ corresponds to contracting clusters and $Q>1/2$ to expanding clusters.
The velocity dispersion of a cluster with a general $Q$ could be written as $\sigma^2 =2QGM_{\rm cl}/R_{\rm cl}$, and the distributions described in subsection \ref{subsec:virial equilibrium} will change correspondingly, such that 

\begin{align}
&n_{\rm{exp}}^{1,2}(\tilde E)\approx n_{\rm{eq}}^{1,2}(\tilde E)/(2Q)^{3/2}, \ 
n_{\rm{exp}}^{1,2}(\tilde a)\approx n_{\rm{eq}}^{1,2}(\tilde a)/(2Q)^{3/2}
\end{align}

\noindent
It should be noted that not only the fraction for each type of binaries is changed, but also the integration boundaries, such that the overall number of binaries grows due to the expansion.
The upper cutoff for an expanding cluster is now not the cluster Hill radius, but the Galactic tidal radius. Hence, not only the normalization changes but also the peak and the shape of the distributions. In principle, a more complicated dependence of the velocity dispersion on the radius could be introduced, which will also change the equilibrium stage of the cluster, i.e. the detailed balance between the creation and destruction of wide/soft binaries will no longer be sustained. However, for the cases we examined, and from the comparison to the N-body simulations, we conclude that the deviation from equilibrium is small. 

\section{Results}\label{sec:results}

In the following, we present the distributions as derived from our model, as well as discuss the capture of triple and higher multiplicity systems and the capture of free-floating planets.

Unless stated otherwise, we consider a cluster with a Kroupa mass function \citep{Kroupa2001}, in the range $0.1\leq m \leq 7 \ M_\odot$, following \cite{PeretsKouwenhoven2012}. For the Monte Carlo simulation we perform, we draw the masses from a Kroupa mass function as implemented in \texttt{AMUSE} \citep{PortegiesZwart2013AMUSE} and its utility \cite{nstarman_zenodo}.
In several clusters we consider also planets, their mass is taken to be $1 \ M_J$, and unless we stated otherwise, we consider in these clusters an equal number of stars and planets, and the size of the cluster by the number of stars. 
The radius of the cluster is set as $R_c=0.1 N_\star^{1/3} \ \rm{pc}$ \citep{PeretsKouwenhoven2012}, such that the number density of stars is kept constant even when the number of stars changes. 

\begin{figure}[H]
    \includegraphics[width=1\columnwidth]
    {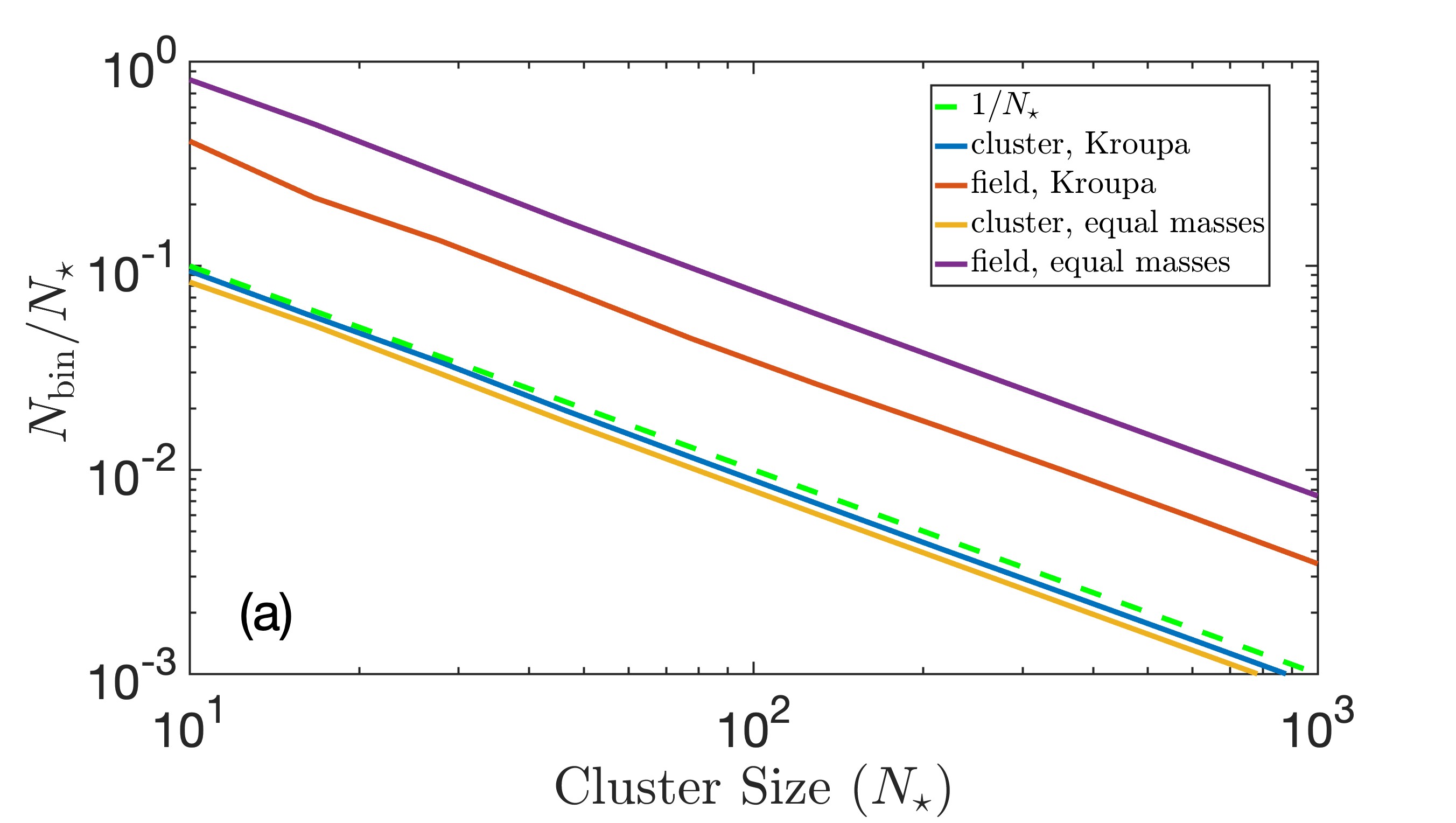}	\includegraphics[width=1\columnwidth]{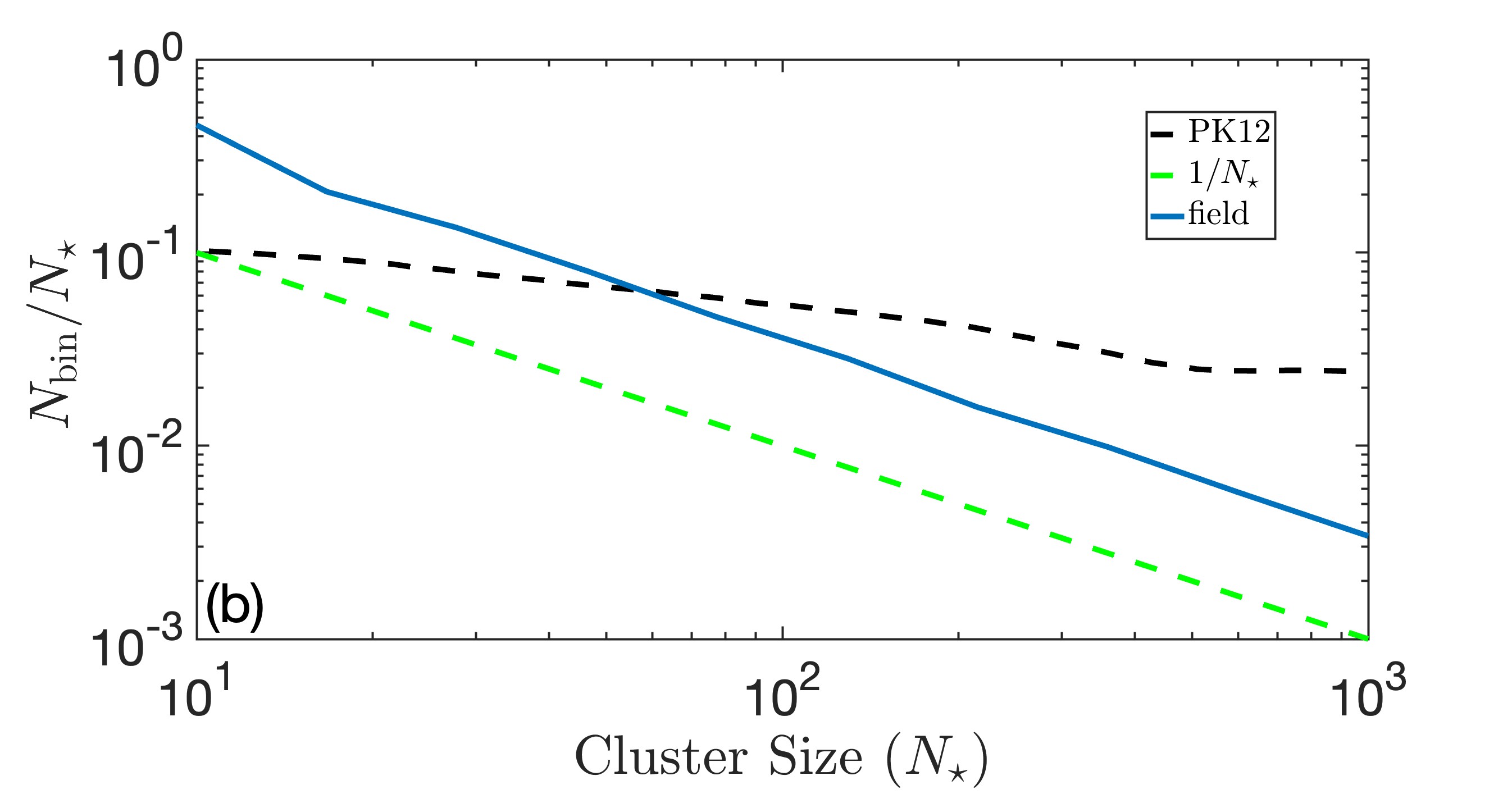}
    \caption{
    Fraction of soft/wide binary systems, as derived from the Monte-Carlo simulation, based on our analytical derivation.
    (a) The fraction of soft binaries in clusters/wide binaries in the field for different choices of mass distribution. 
    (averaged over $100$ Monte-Carlo realizations). (b) The fraction wide of binaries in the field where $N_\star=N_p$ is the number of stars and planets in the systems sampled, in comparison to N-body results (averaged over $50$ Monte-Carlo realizations). 
    }
    \label{fig:fraction}
\end{figure}
\subsection{Wide binary fractions}

In Fig. \ref{fig:fraction} we present the fraction of systems as a function of the cluster size. We present our results for expanding ($Q=3/2$) and equilibrium/non-expanding ($Q=1/2$) clusters, with the corresponding upper semimajor axis cutoffs: the Galactic tidal radius and the Hill radius of the cluster.
As can be seen, the fraction of systems is a monotonically decreasing function of the cluster size, such that the overall scaling agrees well with the freeze-out distribution $1/N_\star$ \citep{MoeckelClarke2011}, with different overall normalization factors for different choices of cutoffs and mass functions. It should be noted that the fraction derived in \cite{PeretsKouwenhoven2012} suggests a more flat dependence on the size of the cluster, which might arise from the specific type of realization of dispersing cluster used in those simulations.

\begin{figure}[H]
\includegraphics[width=1.1\linewidth]{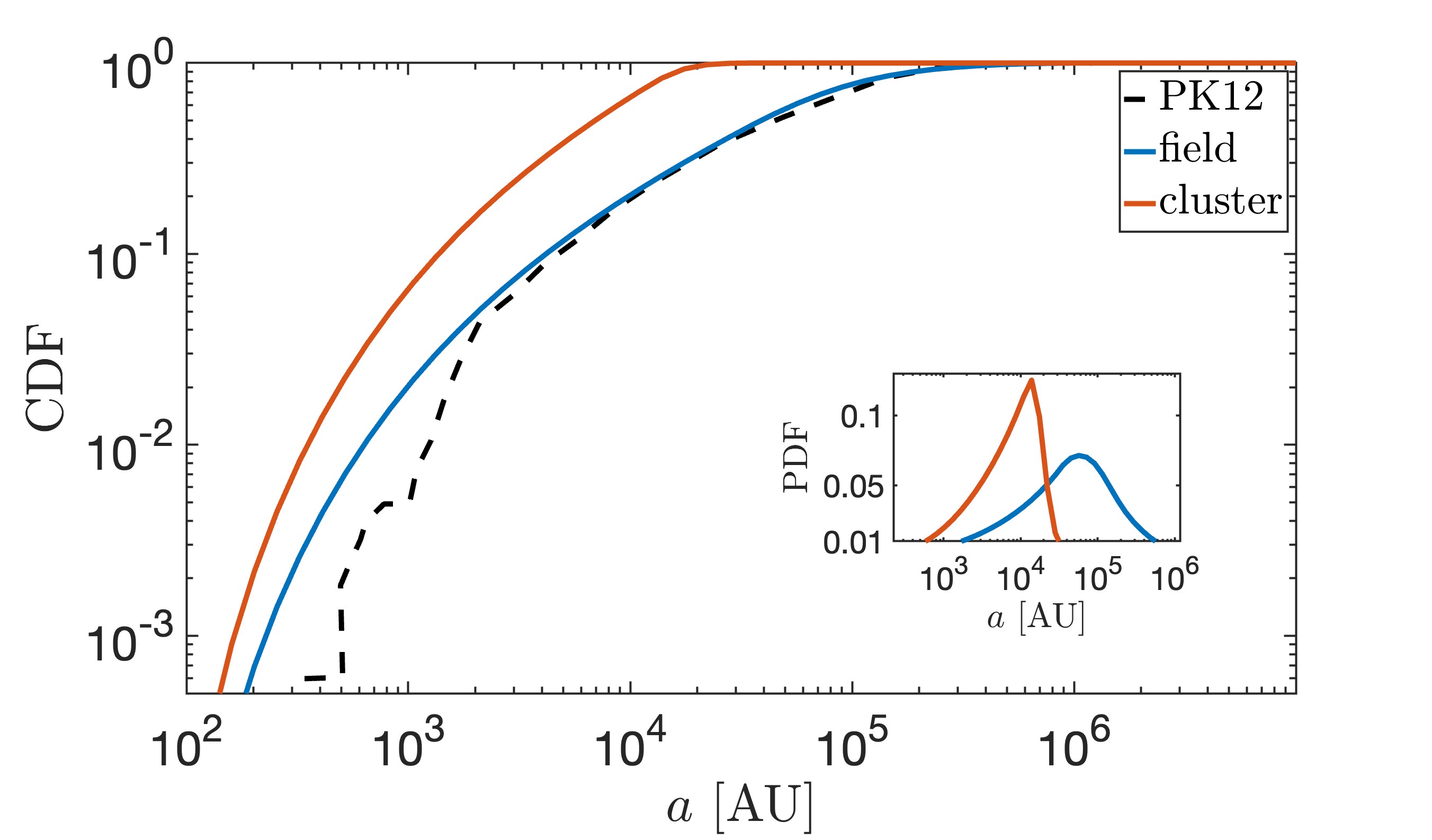}
	\caption{The semi-major axis distribution, as derived from the Monte-Carlo simulation, based on our analytical derivation, averaged over $1000$ realizations, for a cluster containing both stars and planets ($N_\star=N_{p}$), in comparison to results from N-body simulations. }
\label{fig:na}
\end{figure}

\subsection{The properties of wide binaries} 
\subsubsection{Semi-major axis distribution}
In Fig. \ref{fig:na} we present the expected semi-major axis distribution of soft and wide binaries.
The lower cutoff is determined by the transition between soft and hard binaries, i.e. at $a_{\rm SH} = Gm_1m_2/2\sigma^2$ and the upper cutoff is the galactic tidal radius ($R_{\rm tidal,g}$) or the Hill radius of the cluster ($R_{\rm Hill,c}$) correspondingly.
The distribution peaks around $a\sim 10^5  \rm{AU}$ for the galactic tidal cutoff in the field, and at $a\sim 10^4 \rm{AU}$ for the Hill radius cutoff in clusters. nevertheless, the fraction of binaries shows a change by a factor of at most 2-3 over a wide range of semi-major axis (few times $10^3-10^4$ in clusters and few times $10^4-10^5$ AU in the field).  

The expansion of the cluster changes the peak of the distribution, such that in expanding clusters the distribution peaks for higher separations, as expected given the larger allowed upper limit for the separation.   Our results for the distribution are in good agreement with the N-body results presented in \cite{PeretsKouwenhoven2012}, besides the small fraction of lower separations binaries ($<1000$ AU), but at this regime, the small-statistics in the N-body results affect the apparent distribution.

\begin{figure}[H]
\includegraphics[width=1.1\linewidth]{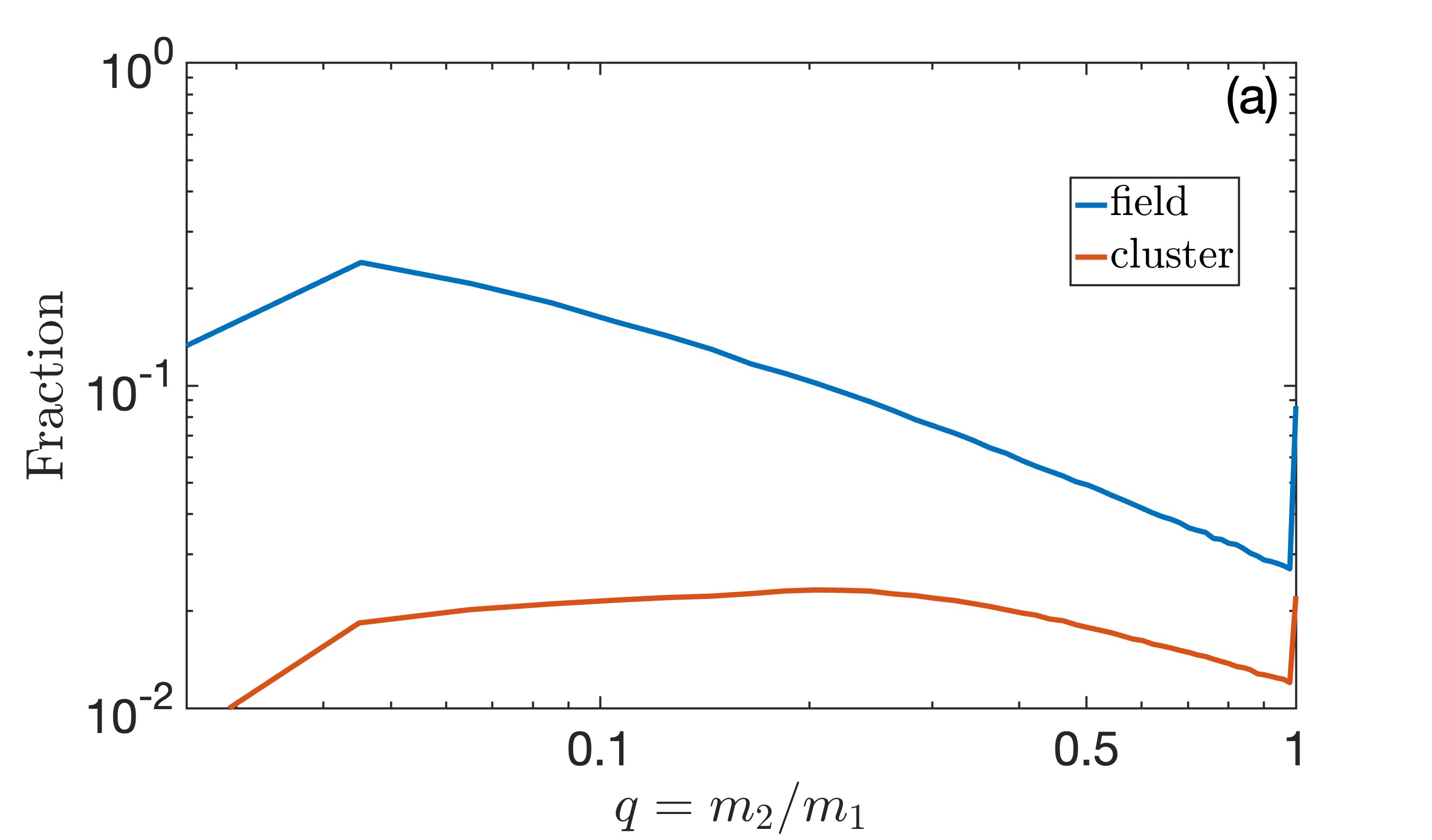}
	\includegraphics[width=1.1\linewidth]{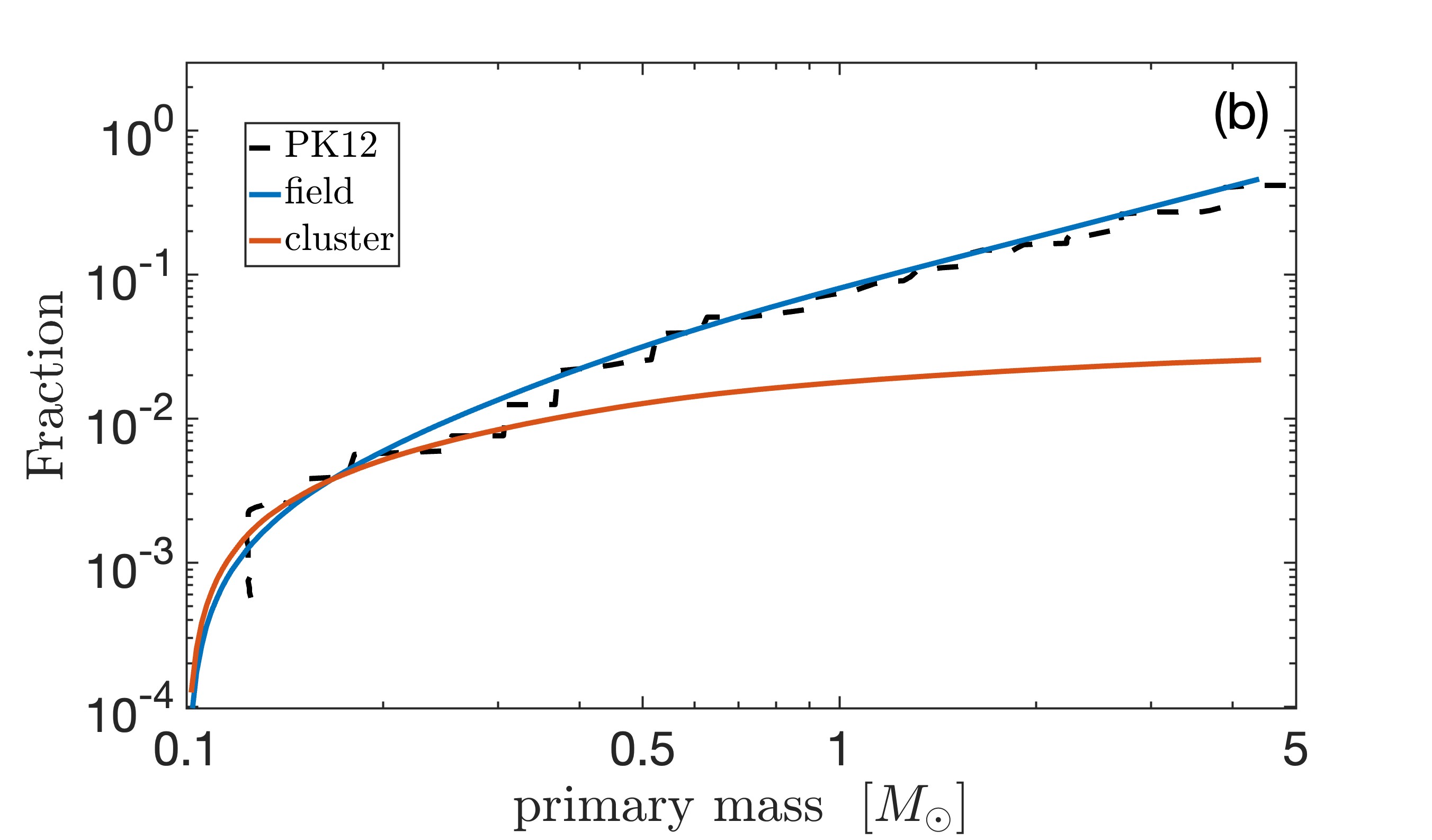}
	\caption{
	The binary fraction for different masses, as derived from an averaging over 1000 runs of the Monte-Carlo simulation, for a cluster containing 100 stars, drawn from a Kroupa mass function, and an equal number of planets, in comparison to N-body simulations. (a) The fraction of binaries as a function of the mass ratio, for stellar binaries only.
	(b) The fraction of binaries as a function of the primary mass.
	}
\label{fig:mass fraction}
\end{figure}

\subsubsection{Masses and mass ratios}
In Fig. \ref{fig:mass fraction}, we present the distribution of captured-formed binaries as a function of mass. We present both the mass ratio distribution (Subfig. a) and 
the primary mass distribution (Subfig. b). As expected from eq. \ref{eq:Fraction}, more massive stars are more likely to reside in binaries, and constitute primaries.
Our results are in excellent agreement with the N-body results presented in \cite{PeretsKouwenhoven2012}.

\subsubsection{Eccentricites \& Inclinations}
In thermalized clusters, binaries are generally expected to show a thermal eccentricity distribution \cite[e.g][]{Ambartsumian1937}, which would then naturally be expected for the wide binaries, which is also consistent with N-body results \citep{PeretsKouwenhoven2012}.
However, different cutoffs of the phase space might lead to deviations from thermal distribution (Rozner \& Perets 2023, in prep.). Observationally, it was found that the eccentricity distribution could slightly deviate from thermal distribution \cite{Raghavan2010,MoeStefano2017,Tokovinin2020}, with twin (same mass) wide binaries are particularly eccentric \citep{Hwang2022twins,Hwang2022}. However, the origin for the latter observation is still not understood, and the wide-binaries formation models explored here are not expected to produce such eccentric "twins".
In populations of non-equal masses, the eccentricity distribution might show some mass dependence, however, here we only explore the distributions of semi-major axes and binary mass function and leave further discussion of eccentricities to later studies. 


Since dynamical captures are generally random, the distribution of the orientations of the formed wide binaries is expected to be randomized. For binaries, this would suggest a random spin-orbit inclination for dynamically-formed wide binaries. 
It was found that binaries with small separations tend to be spin-orbit aligned, while for wide binaries the alignment appears to be random \citep{Hale1994}. However, later studies showed that it might be more complicated, and some correlations may exist (e.g. \citealp{JustesenAlbrecht1994}).
After their formation, wide binaries undergo interaction with external perturbers, that can change their eccentricity and inclination distribution. For triples (see next section), this would suggest a random relative inclination between the inner and the outer orbits of wide triples (and quadruples), which has potential implications for secular processes to play an important role in the evolution, as we discuss below. It should be noted that the eccentricity, as well as inclination, could also be affected by the long-term secular evolution due to the Galactic tide (but it does not explain the observed superthermal distribution of eccentricities \citep{ModakHamilton2023}. 

\subsection{Wide triples and higher multiplicity systems}
Triple and higher multiplicity systems are also known to be abundant \citep{Tokovinin2008,Raghavan2010}, and in particular, most of the massive O/B stars are observed to be part of triples or higher multiplicity systems \citep{Sana2012,MoeStefano2017}. 

Wide triple (and higher multiplicity) systems can also form through the dynamical capture mechanisms discussed above, through two possible channels: (I)  Since primordial binaries are frequent, as discussed above, such systems can dynamically capture additional wide companions, similar to capture by single stars; a hard binary can capture a third distant companion (forming a triple), or another hard binary (forming a quadruple), and this could even result in even higher multiplicity systems if primordial triples etc. are considered. (II) A wide binary formed through dynamical capture, can then consequently  capture additional wide companions. 

The first channel describes the formation of a hierarchical triple in which  $a_{\rm in}\ll a_{\rm out}$, while in the second channel, the capture of an additional companion to an already wide binary might give rise to more comparable inner and outer separations, and even  $a_{\rm in}/a_{\rm out}\sim 1$. In the latter case, the newly formed systems might not be stable and would disrupt over a few dynamical timescales, ejecting one of the stellar components. Here we will briefly discuss each of these channels. 

In the first capture scenario, the initial binary could be thought of as a single object with mass $M_{\rm in}=m_1+m_2$ that captures another object with mass $m_3$, and then the calculation described in section \ref{sec: wide binaries distribution} could be reiterated directly, with the appropriate modifications,

\small
\begin{align}
&N_{\rm tri}^{12,3}(R_{\rm cut})\approx 
\frac{(2\pi)^{3/2}G^{3/2}n_{12}n_3 \bar m^{3/2}}{12\sigma^3}R_{\rm cluster}^3R_{\rm cut}^{3/2}-
\\
\nonumber
&-
\frac{N_1N_2 \pi^{3/2}}{12}\left(\frac{(m_1+m_2)m_3}{M_{\rm cl}^2}\right)^{3/2}
\end{align}
\normalsize

In this scenario, given the known high fractions of binary systems, the capture of a binary system by another binary system is highly likely. Therefore quadruple configurations of 2+2 (two close binary systems orbiting each other at a wide orbit) should be frequent, and secular effects may then later drive the inner binaries into compact configurations through Kozai-cycles and tidal friction. Observations suggest that such quadruple systems are indeed overabundant \citep{Fez+22}, and this scenario might provide a natural explanation. 

The probability for the second channel, i.e. having two consequent captures is $f_{\rm tri,cap}(m_1,m_2,m_3)\approx f(m_1,m_2)\times f(m_2,m_3)$, assuming, without loss of generality, that $m_2$ captures $m_3$ and that two consequent captures are independent. In general, consequent captures could lead to the formation of high multiplicity systems, but with lower probability. In general, the probability of capturing the n-th object with mass m is given by $f_{\rm nm}=f_m^n$ where $f_m$ is the probability of capturing m just once. These probabilities would be somewhat overestimated, in cases where the formed systems are not hierarchical, i.e. considering appropriate stability criteria, (e.g.  \citealp{Valtonen2008}) would be destabilized and lost, lowering the fractions.

\subsection{Contributions to the field population}

The field population consists of contributions from many dispersed clusters. Here we will account for their weighted contributions.
To compare our results with observations, we restrict the mass function from which we sample to the range $[0.1, 3] \ M_\sun$, and the possible separations to be in the range of $100 \ \rm{AU}-1 \ \rm{pc}$, based on the observational results from \cite{ElBadry2021} and references therein. We then sample clusters/associations with $N_\star=10-10^5$ stars from a power-law distribution $dN/dN_\star \propto N_\star^{-2}$, following \cite{LadaLada2003}. We consider for these purposes only wide binaries in the field originated from dispersed clusters. 

\begin{figure}[H]
\includegraphics[width=1.1\linewidth]{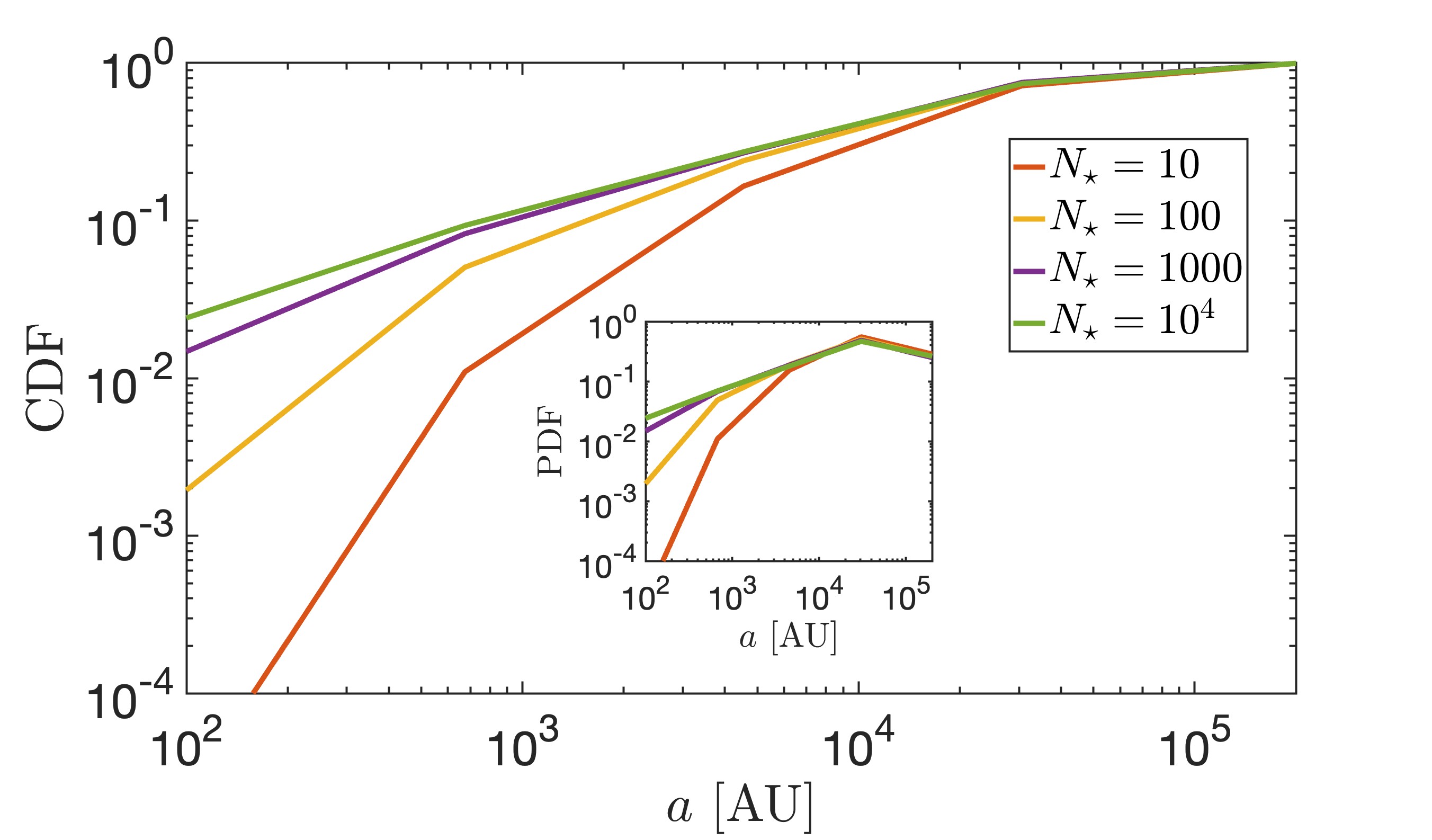}
	\caption{The semi-major axis distribution, as derived from the Monte-Carlo simulation, based on our analytical derivation, for clusters in different sizes. }
\label{fig:mixed}
\end{figure}

 In Fig. \ref{fig:mixed}, we present the separation distribution in different cluster sizes, and the integrated distribution for clusters with $10-10^5$ stars, sampled from $dN/dN_\star\propto N_\star^{-2}$.
The separation distribution is affected by the size of the cluster twice: since $\sigma\propto N_\star^{1/3}$, the overall number density per separation scales as $n_{1,2}(a)\propto N_\star^{-1}$, and the lower separation cutoff scales as $a_{\rm SH}\propto N_\star^{-2/3}$. Hence, the contribution from larger clusters shifts the distribution towards lower separations. Since the powerlaw we chose for the number of stars is negative, the overall distribution is dominated by smaller clusters/associations and our choice of the lower cutoff for the number of stars in these. 

\section{Discussion}\label{sec:discussion}
In the following, we discuss our results for the distributions of wide/soft binaries, as well as discuss the formation of wide triples and higher multiplicity systems and their implications. Some of these were already discussed in \cite{PeretsKouwenhoven2012}; we briefly summarize them, but also new issues and implications. 

\subsection{Comparison of analytic and N-body results}
As discussed above, overall we can reproduce analytically the N-body results presented by \cite{PeretsKouwenhoven2012}, with an excellent agreement, apart from the overall fraction of binaries, in which we find a steeper slope. However, our results for the fractions are in agreement with other N-body models which studied the freeze-out distribution of soft binaries in clusters  \citep{MoeckelClarke2011}.

\subsection{Comparison with observations}

Observational searches of wide binaries usually rely on probabilistic arguments to determine if the binary components are bound, given the uncertainties and/or unknowns in the inferred orbital elements. The era of GAIA data has dramatically expanded the sample of binaries in general and wide binary candidates in particular \citep{GAIA2016,GAIA2018,GAIA2021}, and several studies explored the frequency and distribution of such  binaries  (e.g. \citealp{Oh2017,El-badryRix2018,ElBadry2019,
HartmanLepine2020,Tian2020,
ElBadry2021}). 

The observational wide binary distribution peaks around the separation of $10^3-10^4 \ \rm{AU}$ (\citealp{And+17,ElBadry2021} and references therein), while the distribution we find here peaks at $10^4 \ \rm{AU}$ in clusters and a few times higher in the field. This could be explained by observational biases, the inclusion of cluster binaries in some surveys, the choice of the lower cutoff for the smallest clusters that contribute to the field population, and/or the contribution from other binary-formation channels at smaller separations. In addition, over time wide binaries can experience encounters with field stars. Given the typical velocity dispersion in the field, the widest binaries would evaporate and be disrupted, leaving behind wide binaries of smaller separations, as well as more realistic profiles for the densities of the clusters (e.g. \citealp{BinneyTremaine2010,JiangTremaine2010,MichaelyPerets2016}).

In the equilibrium distribution explored here, The overall wide binary fraction is dominated by the contributions from small clusters/associations, as the cluster-size scales as $dN/dN_\star\propto N_\star^{-2}$, 
each cluster has $N_\star$ stars, such that the overall dependence scales as $1/N_\star\times $binary fraction.
Hence, the lower size cutoff plays an important role in determining the final field properties. For the overall fraction, we consider stellar masses in the range $0.1-7 \ M_\odot$ and clusters in sizes $N_\star=10-10^3$. The total wide binaries fraction for these masses as derived from our model is  $0.2$ for $N_{\star, \rm{min}}\approx 30$; and  $0.006$ for $N_{\star, \rm{min}}=100$.
The observational wide binary fraction ($a>10^3 \ \rm{AU}$) was found to be  $0.115$ \citep{Raghavan2010,MoeckelClarke2011}. \cite{Hwang2021} found a fraction of $0.071$  within $100 \ \rm{pc}$ for binaries with separations larger than $1000 \ \rm{AU}$. Later studies considering a wide range of observations \cite[][and references therein]{MoeStefano2017} found a wide-binary fraction of $0.15$ for 
periods $\log P(\rm{days}) =6.1-7.4$ (corresponding to separations of $\sim 300-2000 \ \rm{AU}$ for solar mass stars). 

\cite{IgoshevPerets2019} discussed the expected wide binary fractions for different mass groups. They studied the binarity of OB stars and found a wide multiplicity rate of $0.091$ for sufficiently luminous secondary stars (G dwarfs or higher) with separations larger than $10^3 \ \rm{AU}$. Extrapolating for lower mass secondaries (assuming a Kroupa mass function for the secondaries), they infer an overall wide binary fraction of $0.27$. 

Setting $N_{\star,\rm{min}}=30$, 
we find a fraction of $0.1$ for G dwarfs (or higher masses) captured by OB-5 stars (cut for our case for $4.36\leq m\leq 7 \ M_\odot$) and a total fraction of $0.5$ for OB captures. These fractions are roughly consistent with the fractions inferred by \cite{IgoshevPerets2019} mentioned above. 

Overall, the fractions we derived are in the expected observational range but are sensitive to the choice for the lower cutoff of clusters'/stellar associations' contribution to field stars. Direct comparison, is difficult for wide binaries of low-mass stars which are long-lived and can be significantly affected by field perturbations and evaporation over time, in particular at larger separations. In that context, short-lived massive stars provide a better direct probe for the pure formation of wide binaries, not affected by long-term dynamical evolution.

As we discussed earlier, the overall binary fraction in clusters scales with $N_\star^{-1}$. \cite{Hwang2021} (and references therein) showed that clusters with lower metallicities, which might indicate higher masses, correspond to larger fractions of wide binaries, which might point on a general agreement with our results. Further comparison with these results is beyond the scope of this paper.

\subsection{Implications for the dynamical evolution of stellar binaries and higher multiplicity systems}

\subsubsection{Collisional field dynamics}
As briefly mentioned in the introduction, wide binaries/triples could serve as catalysts for the collisional dynamics of field stars where flyby perturbations of wide binaries/triples, excite their eccentricities to the point where the components strongly interact and may produce compact binaries, explosive transients and gravitational wave sources \citep{KaibRaymond2014,MichaelyPerets2019,MichaelyPerets2020,GrishinPerets2022,MichaelyNaoz2022}. Our results suggest that dynamical formation of wide binaries and triples could be quite frequent and can occur Myrs after the initial formation of the stellar components. 
In principle, stellar evolutionary scenarios could easily disrupt primordial wide binaries, e.g. due to even low natal kicks and/or Blaauw (prompt mass-loss) kicks (e.g. \citealp{Balaauw1961,HansenPhinney1997,IgoshevPerets2019} and references inside),
in particular for massive stars and their remnant black holes or neutron stars, known to experience prompt mass-loss/natal kicks following core-collapse supernova explosions. One would then think of excluding the possibility of wide BH/NS binaries, and their participation in collisional field dynamics \citep[but see][]{Rav+22}. However, our results provide a channel for the existence of such BH/NS wide binaries, since the dynamical capture, which can occur already after these remnants lost their companions, could allow them to acquire wide companions, which could then indeed play an important role in their evolution through collisional dynamics in the field. In particular, given their relatively large masses, NS, and, to a much higher extent BHs are very likely to capture companions, given our finding of the mass dependence. 

\subsubsection{Secular evolution}

Triple secular and quasi-secular dynamics were suggested to catalyze a variety of phenomena, in particular when the inner and outer orbits are significantly misaligned. Such processes include the formation of short-period binaries (e.g. \citealp{MazehShaham1979,Kiseleva1998}), 
mergers, 
gravitational waves merger sources (e.g. \citealp{Blaes2002,Antognini2014,Antonini2017,LiuLai2018,MichaelyPerets2020}), type Ia supernovae (e.g. \citealp{KatzDong2012,Thompson2011}), blue stragglers \citep{Perets2009Blue_stragglers}
and many others (see for a detailed review of many of these in \citealp{Naoz2016}), 
and hence understanding their properties and distributions analytically could have significant implications. 

The formation of likely misaligned triple and quadruple systems in the capture scenario could naturally provide the necessary conditions for significant secular evolution to take place, and therefore the capture scenario may play a key role in the initial production of secularly evolving systems, and their resulting strong interactions. 

Furthermore, very wide systems are also affected by secular processes triggered by the Galactic tidal field (even for wide binaries, not only higher multiplicity systems).  The dynamics in the galactic fields and other external perturbations were studied extensively (e.g. \citealp{HeislerTremaine1986,JiangTremaine2010,HamiltonRafikov2019a,HamiltonRafikov2019b,HamiltonRafikov2021,GrishinPerets2022} and references therein), and  
it is qualitatively similar to the evolution of quadruple hierarchical systems. Again, our finding on the formation of very wide binaries in the dynamical capture scenario suggests a large number of wide systems, sensitive to galactic tidal secular perturbations exist.  

\subsection{Capture of planets}
 
Free-floating planets (FFPs) are planets that are not bound to any star or brown dwarf. The dynamical interactions between planets are thought to give rise to the frequent ejection of planets, and the production of unbound FFPs \citep{RasioFord96}.

Regardless of the formation channel, 
since the majority of stars are thought to be born and evolve in clusters, so do planets.
In principle, the derivation we used could describe the capture of FFPs by stars or other FFPs. \cite{PeretsKouwenhoven2012} already suggested that during the cluster dispersion planets could be captured, in a similar process. Here we should only mention the potential caveat that planets might not be expected to thermalize with the stars, due to the extreme mass ratio with respect to stars, which will change our distribution assumptions. Nevertheless, following their ejection the velocities of FFPs should initially follow the overall velocity distributions of their original host stars, and might not have time to significantly change before their host cluster disperses.

\section{Summary}\label{sec:summary}

In this paper, we analytically explored the formation of soft binaries in clusters and wide binaries in the field, and derived their properties. We show that our analytic results well reproduce detailed N-body simulations, and can be generally used to derive the properties of dynamically-capture wide binaries in any environment. We find that the capture formation can potentially explain the origin of most of the observed wide binaries, 
and can give rise to higher multiplicity systems, and possibly explain the observed overabundance of wide 2+2 quadruple systems.  
Soft and wide binaries and higher multiplicity systems are highly sensitive to collisional field dynamical processes, as well as secular dynamical processes, all of which may give rise to the formation of compact binaries and/or mergers and transient phenomena. Therefore, the understanding of the formation of wide systems and their properties is of significant importance for the evolution of stars and the production of explosive transients. 

\section*{Acknowledgements}

MR acknowledges the generous support of Azrieli fellowship.




\bibliographystyle{aasjournal}
\bibliography{example.bbl}








\end{document}